\documentclass[%
aip,
amsmath,amssymb,
reprint,%
]{revtex4-1}

\usepackage{graphicx}
\usepackage{dcolumn}
\usepackage{bm}

\usepackage[utf8]{inputenc}
\usepackage[T1]{fontenc}
\usepackage{mathptmx}
\usepackage{etoolbox}
\usepackage{gensymb}
\usepackage{color}
\usepackage{ulem}

\usepackage{soul}

\usepackage{hyperref}
\hypersetup{hypertex=true,
	colorlinks=true,
	linkcolor=blue,
	anchorcolor=blue,
	citecolor=blue}

\begin{document}
\preprint{AIP/123-QED}

\title{Visualization of Photonic Band Structures  via Far-field Measurements in SiN$_x$ Photonic Crystal Slabs}

\author{Wenze Lan}
\affiliation{Beijing National Laboratory for Condensed Matter Physics, Institute of Physics, Chinese Academy of Sciences, Beijing 100190, China}
\affiliation
{School of Physical Sciences, CAS Key Laboratory of Vacuum Physics, University of Chinese Academy of Sciences, Beijing 100190, China}

\author{Peng Fu}
\affiliation{Beijing National Laboratory for Condensed Matter Physics, Institute of Physics, Chinese Academy of Sciences, Beijing 100190, China}
\affiliation
{School of Physical Sciences, CAS Key Laboratory of Vacuum Physics, University of Chinese Academy of Sciences, Beijing 100190, China}

\author{Chang-Yin Ji}
\affiliation
{Centre for Quantum Physics, Key Laboratory of Advanced Optoelectronic Quantum Architecture and Measurement (MOE), School of Physics, Beijing Institute of Technology, Beijing, 100081, China}
\affiliation
{Beijing Key Lab of Nanophotonics \& Ultrafine Optoelectronic Systems, School of Physics, Beijing Institute of Technology, Beijing, 100081, China}

\author{Gang Wang}
\affiliation
{Centre for Quantum Physics, Key Laboratory of Advanced Optoelectronic Quantum Architecture and Measurement (MOE), School of Physics, Beijing Institute of Technology, Beijing, 100081, China}
\affiliation
{Beijing Key Lab of Nanophotonics \& Ultrafine Optoelectronic Systems, School of Physics, Beijing Institute of Technology, Beijing, 100081, China}

\author{Yugui Yao}
\affiliation
{Centre for Quantum Physics, Key Laboratory of Advanced Optoelectronic Quantum Architecture and Measurement (MOE), School of Physics, Beijing Institute of Technology, Beijing, 100081, China}
\affiliation
{Beijing Key Lab of Nanophotonics \& Ultrafine Optoelectronic Systems, School of Physics, Beijing Institute of Technology, Beijing, 100081, China}

\author{Changzhi Gu}
\affiliation{Beijing National Laboratory for Condensed Matter Physics, Institute of Physics, Chinese Academy of Sciences, Beijing 100190, China}
\affiliation
{School of Physical Sciences, CAS Key Laboratory of Vacuum Physics, University of Chinese Academy of Sciences, Beijing 100190, China}

\author{Baoli Liu}
\email{blliu@iphy.ac.cn}
\affiliation{Beijing National Laboratory for Condensed Matter Physics, Institute of Physics, Chinese Academy of Sciences, Beijing 100190, China}
\affiliation
{CAS Center for Excellence in Topological Quantum Computation, CAS Key Laboratory of Vacuum Physics, University of Chinese Academy of Sciences, Beijing 100190, China}
\affiliation
{Songshan Lake Materials Laboratory, Dongguan, Guangdong 523808, China}

\date{\today}

\begin{abstract}
The band structures of the photonic crystal slabs play a significant role in manipulating the flow of light and predicting exotic physics in photonics. In this letter, we show that the key features of photonic band structures can be achieved experimentally by the polarization- and momentum-resolved photoluminescence spectroscopy utilizing the light emission properties of SiN$_x$. The two-dimensional spectra clearly reveal the energy-momentum dispersion of band structures which is in perfect agreement with the simulation results. The isofrequency contours can be measured easily by adding a bandpass filter with a desired photon energy. Furthermore, it is convenient to observe clearly and directly the optical singularity - the optical bound states in the continuum featured by dark point in three-dimensional photoluminescence spectra. The polarization-resolved isofrequency contours clearly show that this dark point is the center of an azimuthally polarized vortex. Finally, the helical topological edge states can be easily observed in photonic topological insulators with deformed hexagonal lattices. Our work provides a simple and effective approach for exploring topological photonics and other intriguing phenomena hidden in the photonic crystal slabs. 
\end{abstract}

\maketitle

The photonic crystals offer the huge opportunities to manipulate the flow of light\cite{joannopoulos1997,2011Joannopoulos}. As a class of photonic crystals, the photonic crystal (PhC) slabs can be fabricated easily by the state-of-the-art semiconductor nanofabrication process and promotes significantly the rapid developments of the fundamental research and device application in photonics, especially topological photonics\cite{Lu2014topological,Ozawa2019}. The numerous intriguing physical phenomena hidden in photonic crystals and advanced multi-functional photonic devices are revealed experimentally and predicted theoretically such as the unidirectional transport of light robust against scattering losses and arbitrarily large disorder\cite{Wang2009,Poo2011}, supercollimation\cite{rakich2006}, bound states in the continuum (BICs)\cite{2008Marinica,2011Plotnik,2013Hsu,Hsu2016bound}, large-area surface-emitting lasers\cite{matsubara2008,hirose2014,yang2022topological}, non-Hermitian physics\cite{Zhen2015spawning,zhou2018observation}, the photonic analogues of the quantum (anomalous) Hall effect\cite{Haldane2008,Wang2009}. All above-mentioned phenomena result from the special band structures of the photonic crystals. For instance, the supercollimation relies on the flatness isofrequency contours; the BICs singularity point and surface-emitting lasers are significantly related to the strongly confined electric fields and zero group velocities of electromagnetic waves.

The widely used technique to investigate the band structures of the photonic crystal slabs is the white light reflection/transmission spectra through the analysis of Fano features exhibited in spectra\cite{2002FanSH,2018ZhangYW}. The isofrequency contours can be also obtained by passive reflection or transmission modes based on 4f momentum space imaging system\cite{2018ZhangYW,2021ZhangYW}.  Meanwhile, the technique of photon scattering was developed to directly observe the isofrequency contours\cite{ShiL2010,2016Regan}. The advantage of these techniques is that all of them are suitable for the most photonic crystals. As a complementary technology, the momentum-resolved photoluminescence (PL) spectroscopy can be explored  to access the photonic band structures via the far-field measurements of the excitonic emission\cite{2019HuangWZ} or enhanced fluorescence\cite{Zhen2013,2020Seo}  due to cavity-enhanced light emission. That the PL peak at a certain k point directly reveals a photonic state in the band structures is the superiority of far-field measurements although this technique relies on the light emission properties of the photonic crystals or the extra emitters. 

The SiN$_x$ thin films, as the ideal optical materials with enough index contrast (optical index n$\approx$2.2), are extensively  explored to fabricate the various photonic crystals \cite{2012Lee,2013Hsu,Zhen2015spawning,zhou2018observation,2019LiuWZ,Liu2020z2} through the mature nanofabrication technology in semiconductor industry.  It is worth noticing that the SiN$_x$ exhibits the light emission properties with relatively larger bandwidths of more than $\sim$ 400 nm (See Fig.S1 in Supplementary Materials) induced by the quantum-size effects of Si nanoparticles\cite{WangYQ2003}, SiN$_x$ matrix\cite{Kistner2011} or defect-related states\cite{wang2007photoluminescence} during fabrication.

Here, fully utilizing the light emission properties of the SiN$_x$ materials\cite{Dal2006,wang2007photoluminescence,Kistner2011,Amosov2022}, we investigate experimentally the photonic band structures in the two-dimensional free-standing SiN$_x$ PhC slabs by the polarization- and momentum-resolved photoluminescence spectroscopy based on the 4f optical system.  The detailed photonic band structures of PhC slabs can be visualized and achieved : 1) energy-momentum dispersion with the accurate energy level of photonic states at any direction in momentum space; 2) isofrequency contours at any energy level without changing the optical setup; 3) the degree of confinement and the distribution of the confined electromagnetic modes in photonic crystal slab through the analysis of PL intensities; 4) polarization of photonic states. More important, except for the fundamental information of photonic band structures, it is also convenient to directly observe the fascinating phenomena of PhC slabs, for instance, the optical bound states in the continuum (BICs) featured by dark point in dispersion relation curves and polarization vortex in isofrequency contours; the helical topological edge states in the photonic topological insulator. Our work provides a simple and effective approach for exploring the intriguing physical phenomena hidden in the photonic crystal slabs.

\begin{figure}[t]
	\centering
	\includegraphics[width=0.5\textwidth]{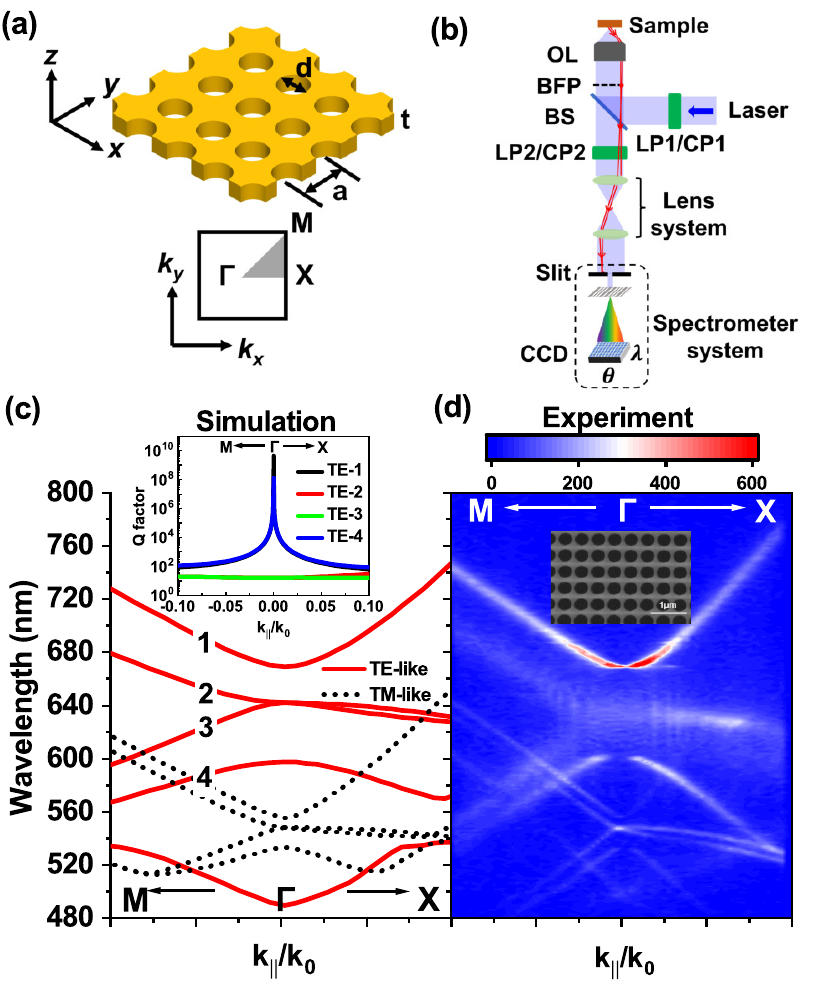}\\
	\caption{(a) Schematic of the SiN$_x$ freestanding photonic crystal slab and its first Brillouin zone with high symmetry points $\Gamma$, M and X. The lattice constants, hole diameter, and thickness are represented by the symbols $a$, $d$, and $t$, respectively. (b) Illustration of the home-built 4f optical system. OL: objective lens. BFP: back focal plane. LP: linear polarizer. CP: circular polarizer. BS: beamsplitter. (c) Simulated TE-like and TM-like photonic bands. The inset shows the simulated quality factors of four labeled TE-like modes. (d) The measured photonic dispersion is represented by momentum-resolved photoluminescence spectra. The ranges of $k_{\Vert}/k_0$ in band structures in (c) and (d) are from -0.4 to 0.4. The inset shows the scanning electron microscopy image of the sample. Scale bar: 1 $\mu$m.}
	\label{Figure 1}
\end{figure}

To evaluate the abilities of this advanced method, we select firstly the SiN$_x$ membrane to fabricate the two-dimensional photonic crystal. Figure \ref{Figure 1}(a) presents the schematic of the designed photonic crystal slab and its first Brillouin zone with high symmetry points. This PhC slab consists of a square lattice of cylindrical air holes with a radius of 180 nm and a period of 490 nm. Except for the general photonic states, the optical singularity, so-called the symmetry-protected BICs ($\Gamma$-BICs), will be created because the mirror symmetry of PhC slab in out-of-plane direction was maintained naturally due to the key feature of a suspending SiN$_x$ membrane, which also can reduce the extra radiation channels\cite{2011Joannopoulos,2019LiuWZ} and allows us to classify the modes of transverse electric (TE)-like and transverse magnetic (TM)-like clearly. Figure \ref{Figure 1}(b) presents the schematic of our home-made 4f optical system, which is based on the optical Fourier transformation method achieved through the optical lenses to obtain the momentum-space information of photonic states of band structures \cite{2012Wagner,2021ZhangYW}(See Optical Characterization in Supplementary Materials). 

We first calculate the band structures of the designed PhC slab. Figure \ref{Figure 1}(c) presents the simulated TE-like (red solid line) and TM-like (black dash line) photonic bands along the $\Gamma$X and $\Gamma$M directions. For the sake of simplicity, the physical properties of the four TE-modes as labelled in Figure \ref{Figure 1}(c) will be focused on emphatically in the following main text. It is deserved to note that the TE-1 and TE-4 photonic bands exhibit the parabolic profiles with extreme points of photonic energies located at $\Gamma$-points with the corresponding wavelengths $\sim$670 nm and $\sim$590 nm, respectively, in which the group velocities of electromagnetic waves for TE-1 and TE-4 modes are equal to zero through the equation $v_{g}=\nabla_{\bold{k}}\omega(\bold{k})$. It is expected that the electromagnetic waves for TE-1 and TE-4 modes are completely confined in the PhC slab at $\Gamma$-points, which will result in the high quality factor when this PhC slab acts as an optical cavity. The calculated quality factors of four labeled TE bands are shown in the inset of Figure \ref{Figure 1}(c). Both the TE-1 and TE-4 bands possess the extremely high quality factors at $\Gamma$-points which are the unique signature of BICs\cite{2014ZhenB}. Meanwhile, the quality factors are almost flat and lower in the whole k space for the TE-2 and TE-3 bands. Figure \ref{Figure 1}(d) presents the experimental results of photonic dispersion along the $\Gamma$X and $\Gamma$M directions by performing the measurements of the momentum-resolved photoluminescence spectroscopy on a fabricated PhC slab as shown in the inset of Figure \ref{Figure 1}(d) (see Methods for details of PhC slab fabrication in Supplementary Material). It is clear that the measured photonic dispersion is consistent with the simulation results very well. The whole TE- and TM-modes can be obtained from this two-dimensional PL spectra with a very clean background and without any data post-processing. 

\begin{figure}[t]
	\centering
	\includegraphics[width=0.5\textwidth]{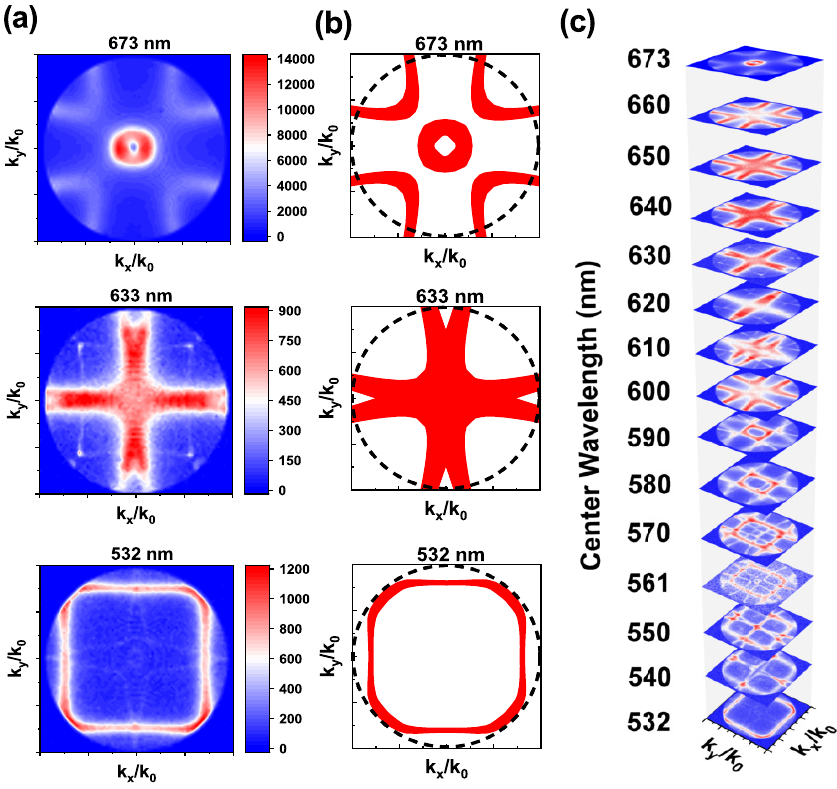}\\
	\caption{(a) The measured and (b) the simulated isofrequency contours of center wavelength at 673 nm, 633 nm, and 532 nm. The simulation results cover the  practical bandwidth of the bandpass filters. The black dashed circle indicates the detection range. (c) The isofrequency contour slices at several central wavelengths. The ranges of $k_x/k_0$ and $k_y/k_0$ in (a), (b) and (c) are both from -0.4 to 0.4.}
	\label{Figure 2}								
\end{figure}

Another more comprehensive understanding of the optical properties of a PhC slab lies in its isofrequency contours which need to be achieved experimentally. Concerning to our approach, only fully opening the slit and setting the grating as a mirror (zero position), the isofrequency contours can be directly imaged through the insertion of a set of bandpass filters on the detection path. The simulated and experimental results of 532 nm, 633 nm (both bandwidths $\sim$4 nm), and 673 nm (bandwidth $\sim$10 nm) are presented in Figure \ref{Figure 2}(a) and (b), respectively. Again, it is obvious the consistency of both results is quite good. By replacing the bandpass filter, we can easily obtain the isofrequency contour at a desired photonic energy as shown in Figure \ref{Figure 2}(c).

\begin{figure}[t]
	\centering
	\includegraphics[width=0.5\textwidth]{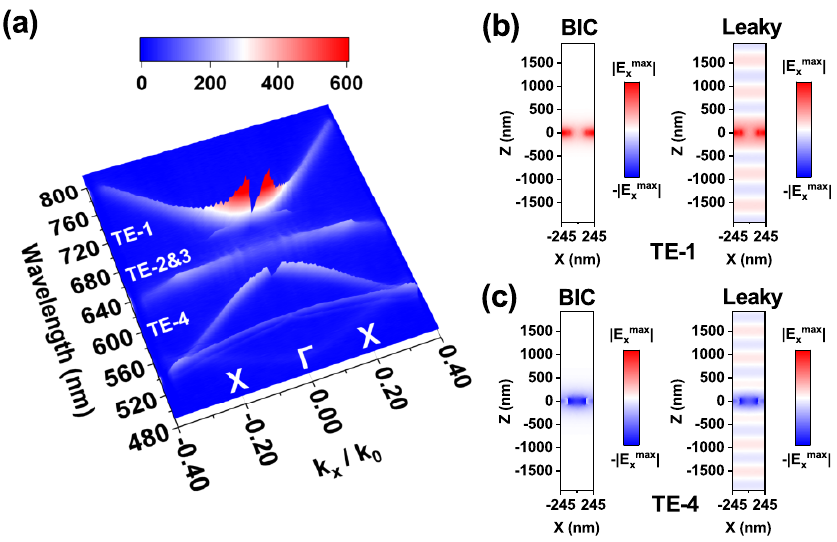}\\
	\caption{(a) The 3D colormaps of photonic bands in $\Gamma$X direction. (b-c) Simulated electric field profiles ($x$ component) of the at-$\Gamma$ BIC (${\bf{k}}_{\Vert}/k_0=(k_x,k_y)/k_0=(0,0)$) and off-$\Gamma$ leaky mode (${\bf{k}}_{\Vert}/k_0=(k_x,k_y)/k_0=(0,0.07)$) sliced on the plane $y=-120$ nm for the TE-1(b) and TE-4(c) bands.}
	\label{Figure 3}
\end{figure}

It is worth noting that our experimental data include the information of PL intensities as presented in Figure \ref{Figure 1}(d). Generally speaking, the presented experimental method belongs to the far-field measurements of leaky modes in photonic crystal slabs. Therefore, the information of PL intensities can reveal how much extent the electromagnetic modes are confined and where the electromagnetic energy is concentrated in photonic crystal slabs. We will elaborate the advantages of our experimental approach through the analysis of PL intensities in photonic dispersion along $\Gamma$X, which can be directly visualized by plotting the experimental data as the 3D colormaps shown in Figure \ref{Figure 3}(a). Both TE-1 and TE-4 bands exhibit the minima of PL intensities (denoted as dark points ) at the $\Gamma$ point corresponding to the wavelength of $\sim$670 nm and $\sim$590 nm, respectively, in photonic dispersion along $\Gamma$X. The  dark points also can be observed in photonic dispersion along $\Gamma$M (data not shown here). The dark points are the distinguished feature of the symmetry-protected BICs with extremely high quality factors as shown in the inset of Figure \ref{Figure 1}(c). Side views of simulated electric field profiles confirm that the BIC mode is completely confined in the photonic crystal slab compared to the off-$\Gamma$ leaky mode as shown in Figure \ref{Figure 3}(b, c). For TE-2 and TE-3 bands, there are no  dark points in photonic dispersions due to the lower quality factors. Furthermore, it is noticed that the PL intensities around the $\Gamma$ point in TE-1 band are higher than that in TE-4 band although the quality factors at $\Gamma$ point are the same for those two bands. From Figure \ref{Figure 3}(b, c), we can see that the energy of electromagnetic modes, which can be represented by the magnitude of the electric field, is concentrated in SiN$_x$ slab (high-$\epsilon$ regions) and air hole (low-$\epsilon$ regions) for TE-1 and TE-4 bands, respectively. In our experiment, the SiN$_x$ slab also plays a role as the light emitter. Therefore, the stronger light-matter interaction in SiN$_x$ slab will result in the enhancement of PL intensities in TE-1 band compared to that in TE-4 band via Purcell effect\cite{Purcell1946}. This well-known feature that where the energy of electromagnetic modes is concentrated in low-$\epsilon$ or high-$\epsilon$ regions\cite{2011Joannopoulos} can not be addressed by the traditional reflection measurements.

\begin{figure}[t]
	\centering
	\includegraphics[width=0.5\textwidth]{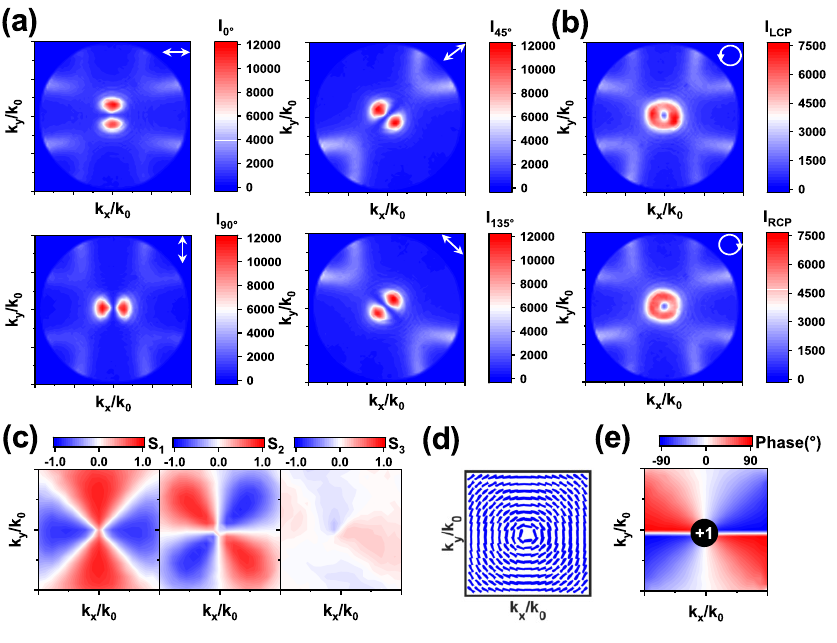}\\
	\caption{(a-b) Polarization-resolved isofrequency contours near the BIC position ($\sim$673 nm) with (a) linearly polarized ($I_{0\degree}$, $I_{45\degree}$, $I_{90\degree}$, $I_{135\degree}$) and (b) circularly polarized detection ($I_{LCP}$, and $I_{RCP}$). (c) Normalized Stokes parameter $S_1$, $S_2$ and $S_3$. (d) Polarization ellipse map around the $\Gamma$ point. (e) The Stokes phase map around the $\Gamma$ point. The ranges of $k_x/k_0$ and $k_y/k_0$ in (a) and (b) are both from -0.4 to 0.4. The ranges of $k_x/k_0$ and $k_y/k_0$ in (c), (d) and (e) are both from -0.1 to 0.1.}
	\label{Figure 4}
\end{figure}

Now, we turn to characterize the optical properties of photonic states around BICs in the TE-1 band via this method. As an optical singularity, BICs is a vortex center characterized by the value of topological charge and polarization distribution of photonic states around BICs\cite{2014ZhenB}, which can be extracted through the measurements of polarization-resolved isofrequency contours and the analysis of the Stokes phase distribution, benefiting from the very closed wavelength of BICs and the center wavelength of the filter. It is noted that the isofrequency contour labeled as $\sim$673 nm in Figure \ref{Figure 2}(a) exhibits a donut shape with a dark zone at the center, which is the typical signature of a vortex beam. Figure \ref{Figure 4}(a) and (b) present the polarization-resolved isofrequency contours. With linearly polarized detection, two lobes are observed as shown in Figure \ref{Figure 4}(a). With rotating the polarizer, the connection direction of the two lobes is still perpendicular to the polarization direction of the linear polarizer. For circularly polarized detection, the isofrequency contours in Figure \ref{Figure 4}(b) still exhibit a donut shape as a result of a coherent superposition of two orthogonally linear polarization. Those experimental observations demonstrate that BICs are the center of the azimuthal polarization vortex\cite{2009ZhanQW,2011Iwahashi,2019ShenYJ} and the photonic states around BICs are linearly polarized. Next, we determine quantitatively the value of topological charge $\textit{q}$ defined as following:
\begin{equation}
    q=\frac{1}{2\pi}\oint_L d\bm k_{\parallel}\cdot\nabla_{\bm k_{\parallel}}\phi(\bm k_{\parallel})
    \label{equation 1}
\end{equation}
where  $L$ is a closed loop around the BICs in the counterclockwise direction, $ \phi(\bm k_{\parallel})$ is the Stokes phase expressed as:
\begin{equation}
   \phi(\bm k_{\parallel})=\frac{1}{2}arg(S_1+iS_2)
   \label{equation 2}
\end{equation}
The Stokes parameters $S_1=\frac{I_{0\degree}-I_{90\degree}}{I_{0\degree}+I_{90\degree}}$, $S_2=\frac{I_{45\degree}-I_{135\degree}}{I_{45\degree}+I_{135\degree}}$ and $S_3=\frac{I_{LCP}-I_{RCP}}{I_{LCP}+I_{RCP}}$ can be obtained through the measurements of six polarization components ($I_0$, $I_{45\degree}$, $I_{90\degree}$, $I_{135\degree}$, $I_{LCP}$, $I_{RCP}$) \cite{gbur2016singular}. In general, the Stokes parameters $S{_1}$ and $S{_2}$ can fully describe the linear polarization characteristics of the measured photonic states. Figure \ref{Figure 4}(c) shows the two-dimensional (2D) maps of $S{_1}$, $S{_2}$ and $S{_3}$ in momentum space around BICs ($\Gamma$ point). Again, it is clear that the photonic states around BICs are almost linearly polarized as shown in Figure \ref{Figure 4}(d). At the $\Gamma$ point of $k_x=k_y=0$, the polarization of photonic states cannot be defined. Now we calculate the Stokes phases in momentum space with the Stokes parameters $S{_1}$ and $S{_2}$ via Eq. (\ref{equation 2}). Figure \ref{Figure 4}(e) presents the 2D map of Stokes phases in momentum space. It is clear that the change of Stokes phase is 2$\pi$ around BICs in the counterclockwise direction. Therefore, combined with the observations of the two spiral arms in equal-thickness interference and the bifurcation pattern in equal-inclination interference (See Figure S2 in Supplementary Material), the topological charge of $q$ = +1 can be deduced through Eq. (\ref{equation 1}), which is consistent with the former reports\cite{2020WangB,mohamed2022controlling,Salerno2022}. Those results demonstrate that the experimental approach presented here can be employed to conveniently and effectively characterize not only the common features such as energy-momentum dispersion but also the optical singularity in photonic crystal slab, for instance, BICs, through the measurements of far-field radiation\cite{2014ZhenB,2020WangB}. 

\begin{figure}[t]
	\centering
	\includegraphics[width=0.5\textwidth]{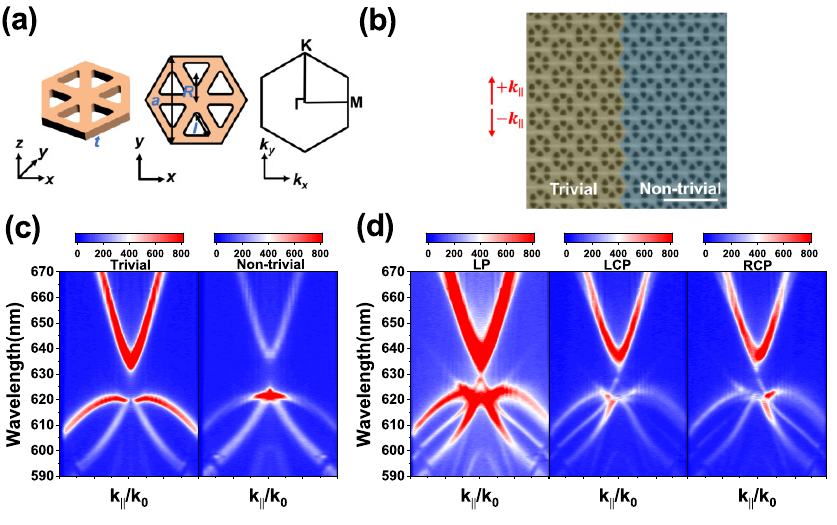}\\
	\caption{(a) Schematic of a graphene-like SiN$_x$ freestanding PhC with a hexagonal lattice of etched triangular air holes and its Brillouin zone. (b) SEM image of the photonic topological insulator. Red arrows indicate the direction of the detected momentum. Scale bar: 1 $\mu$m. (c) The measured photonic dispersions along the $\Gamma$-$K$ directions for the trivial and non-trivial lattices. (d) Edge states dispersions under linearly polarized(LP), left circularly polarized (LCP), and right circularly polarized (RCP) detections. The ranges of $k_{\Vert}/k_0$ in (c) and (d) are from -0.4 to 0.4.}
	\label{Figure 5}
\end{figure}

Finally, we apply this method to explore the helical topological edge states in photonic topological insulators which is the key feature in topological physics. Figure \ref{Figure 5}(a) shows the schematic of the unit cell of graphene-like topological photonic crystals with a hexagonal lattice of etched triangular air holes and its Brillouin zone. The thickness and lattice period of the PhC are $t$=100 nm and $a$=496 nm, respectively. All triangular air holes have a side length of $I$=150 nm and a fillet of 25 nm. $R$ is the distance from the center of the triangular air hole to the center of the unit cell. When $R<a/3$ or $R>a/3$, the lattice is called the trivial lattice or non-trivial lattice\cite{ji2022}, respectively. Figure \ref{Figure 5}(b) presents the SEM image of the photonic topological insulator composed of trivial ($R$=148 nm) and non-trivial ($R$=175.5 nm) lattice regions connected by a zigzag domain wall, which results in the helical topological edge states. The energy-momentum dispersions are measured in trivial and non-trivial lattice regions along the $\Gamma$-$K$ directions as presented in Figure \ref{Figure 5}(c), respectively. There exists the noticeable photonic band gaps between 620nm and 630nm in both lattice regions. When the detection position is put on the zigzag domain wall, the helical topological edge states, which connect the upper and lower branches of photonic bands with almost linear dispersion, can be conspicuously distinguished within the photonic band gap as shown in Figure \ref{Figure 5}(d).

In summary, we present a convenient and effective experimental method to investigate the photonic band structures in dielectric SiN$_x$ photonic crystal slabs combined with the well-known polarization- and momentum-resolved photoluminescence spectroscopy. The detailed photonic band structures such as energy-momentum dispersion, isofrequency contours, etc., can be fully accessed with a very clean background. Except for the general information of photonic band structures, the optical singularity-BICs and the helical topological edge states can be distinguished conveniently and clearly. We believe that this method can be easily exploited by other researchers in the community of photonics, especially topological photonics, and will boost significantly topological photonics.

\vspace{10pt}
See the supplementary material for the details of the numerical simulation, sample fabrication, optical characterization, and extended data for the photoluminescence spectra of SiN$_x$ membrane and interference measurement of vortex beam.

\begin{acknowledgments}
This work was supported by the National Key Research and Development Program of China under Grant No. 2021YFA1400700, the Strategic Priority Research Program of Chinese Academy of Sciences (CAS) under Grant Nos. XDB28000000 and XDB33000000, the National Natural Science Foundation of China under Grant Nos. 11974386, 11904019, 12074033, the Beijing Municipal Science \& Technology Commission, Administrative Commission of Zhongguancun Science Park under Grant No. Z211100004821009, the Key Research Program of Frontier Sciences of CAS under Grant Nos. QYZDJ-SSWSLH042. 
\end{acknowledgments}

\section*{Author Declarations}
\subsection*{Conflict of Interest}
The authors have no conflicts to disclose.


\section*{Data Availability}
The data that support the findings of this study are available from the corresponding author upon reasonable request.

\bibliography{ref}

\end{document}